\documentclass[final,3p,times,8pt]{elsarticle}
\usepackage{amssymb}
\usepackage{lineno}

\usepackage{graphicx}% Include figure files
\usepackage{amsmath}
\usepackage{subfig}
 \def\vector#1{\mbox{\boldmath $#1$}}

\makeatletter
\def\@author#1{\g@addto@macro\elsauthors{\normalsize%
    \def\baselinestretch{1}%
    \upshape\authorsep#1\unskip\textsuperscript{%
      \ifx\@fnmark\@empty\else\unskip\sep\@fnmark\let\sep=,\fi
      \ifx\@corref\@empty\else\unskip\sep\@corref\let\sep=,\fi
      }%
    \def\authorsep{\unskip,\space}%
    \global\let\@fnmark\@empty
    \global\let\@corref\@empty  %% Added
    \global\let\sep\@empty}%
    \@eadauthor={#1}
}
\makeatother

\makeatletter
\def\ps@pprintTitle{%
   \let\@oddhead\@empty
   \let\@evenhead\@empty
   \let\@oddfoot\@empty
   \let\@evenfoot\@oddfoot
}

\begin{document}

\begin{frontmatter}

%% Title, authors and addresses

%% use the tnoteref command within \title for footnotes;
%% use the tnotetext command for the associated footnote;
%% use the fnref command within \author or \address for footnotes;
%% use the fntext command for the associated footnote;
%% use the corref command within \author for corresponding author footnotes;
%% use the cortext command for the associated footnote;
%% use the ead command for the email address,
%% and the form \ead[url] for the home page:
%%
\title{Controlling the dynamics of cloud cavitation bubbles through acoustic feedback}
%% \title{Title\tnoteref{label1}}
%% \tnotetext[label1]{}
\author[Stanford]{Kazuki Maeda\corref{cor1}}
\ead{kemaeda@stanford.edu}
\cortext[cor1]{Corresponding author}

\author[UWM]{Adam D. Maxwell}

%\title{Symmetry reduction of the volume-averaged Navier-Stokes equation for
%simulations of dispersed multiphase flows}

%% use optional labels to link authors explicitly to addresses:
%% \author[label1,label2]{<author name>}
%% \address[label1]{<address>}
%% \address[label2]{<address>}

%\author{Kazuki Maeda$^*$ and Tim Colonius}

\address[Stanford]{Center for Turbulence Research, Stanford University}
\address[UWM]{Department of Urology, University of Washington School of Medicine}

\begin{abstract}
%% Text of abstract
Cloud cavitation causes nontrivial energy concentration and acoustic shielding in liquid, and its control is a long-standing challenge due to complex dynamics of bubble clouds. We present a new framework to study closed-loop control of cavitation through acoustic feedback. While previous approaches used empirical thresholding, we employ model-based state estimation of coherent bubble dynamics based on theory and high-performance computing. Using a pulsed ultrasound setup, we demonstrate set-point control of the pulse repetition frequency (PRF) to modulate acoustic cavitation near a solid target over $O(100)$ s. We identify a quasi-equilibrium correlation between PRF and the bubble dynamics, and an optimal PRF to minimize acoustic shielding of the target. This framework can be readily scaled up by enhanced acoustic sensing and computational power.
\end{abstract}

\end{frontmatter}

%%
%% Start line numbering here if you want
%%
%%\linenumbers

%% main text
Control of cloud cavitation - nucleation of bubble clusters due to rapid fall of the local pressure in liquid - is a long-standing challenge for optimization of medical and hydraulic systems, and sonoluminescence \citep{Plesset55,Plesset77,Crum79,Morch80,Reisman98,Brenner02,Pishchalnikov03,Ando12,Cairos17}.
For example, in extracorporeal ultrasound (US) therapy, the tensile component of high-intensity focused ultrasound (HIFU) can nucleate cavitation bubbles in the human body.
These bubbles violently oscillate and collapse at a submicrosecond time scale to cause damage in surrounding materials \citep{Crum88,Pishchalnikov03,Coussios07,Movahed16} as well as energy shielding of the targets \citep{Pishchalnikov06b,Maeda18a}. The intensity of cavitation can largely fluctuate due to non-equilibrium, stochastic nature of nucleation events, depending on the applied pressure fields \citep{Brotchie09}.
Once nucleated, bubble clouds can persist and be proliferated by subsequent waves that arrive before dissolution \citep{Pishchalnikov11,Maxwell11,Frank15}.
These fascinating bubble dynamics have been quantified through advanced experiments \citep{Bremond06,Ohl15,Cairos17}, although direct observation is limited to specialized setups. In practice, far-field, bubble-scattered acoustic signals are the only observable quantities.
For US-induced acoustic cavitation, open-loop control of the US waveform has been explored to favorably trigger violent collapse of bubble clouds to enhance cavitation erosion \citep{Ikeda06}.
Closed-loop control has been utilized to excite stationary cavitation by modulating incident US waves, such that a stable acoustic feedback is maintained \citep{Thomas05,Hockham10,Sabraoui11,Patel18,Tan19}.
These systems, however, rely on empirical thresholds and lack a quantifiable state estimation of cavitation, as the bubble dynamics are not modeled in the feedback loop. This drawback motivates us to pursue model-based feedback control.

Modeling of cloud cavitation has been extensively explored in the past decades \citep{Brennen14}. A recent effort has identified a scaling parameter that dictates the coherent dynamics of spherical bubble clouds in ultrasound fields, {\it the dynamic cloud interaction parameter}: $B_D=N<R_b>/R_c$, where $N$ and $R_b$ are the total number and the average radius of bubbles in the cloud, $R_c$ is the radius of the cloud, and $<>$ denotes the time-average during periodic oscillations \citep{Maeda19}. The parameter characterizes the structure of bubble clouds and the bubble-induced acoustic fields. In the limit of linear oscillations of bubbles, a static form of such a parameter can be obtained using a mean-field theory \citep{dAgostino89}. $B_D$ was derived from the first-principle hydrodynamic many-body theory and extends to nonlinear dynamics of cavitation bubbles that grow far from equilibrium.
In experiments, correlations have been identified between the energy state of bubble clouds and the bubble-scattered acoustic waves \citep{Maeda18a,Maeda19}. These results are promising for real-time estimation of bubble dynamics through acoustic measurements \citep{Maeda18t}.

In this work, we design a framework to study model-based closed-loop control of cloud cavitation through acoustic feedback, and demonstrate its implementation in a pulsed US system.
\begin{figure}[h!]
\centering
\includegraphics[width=110mm]{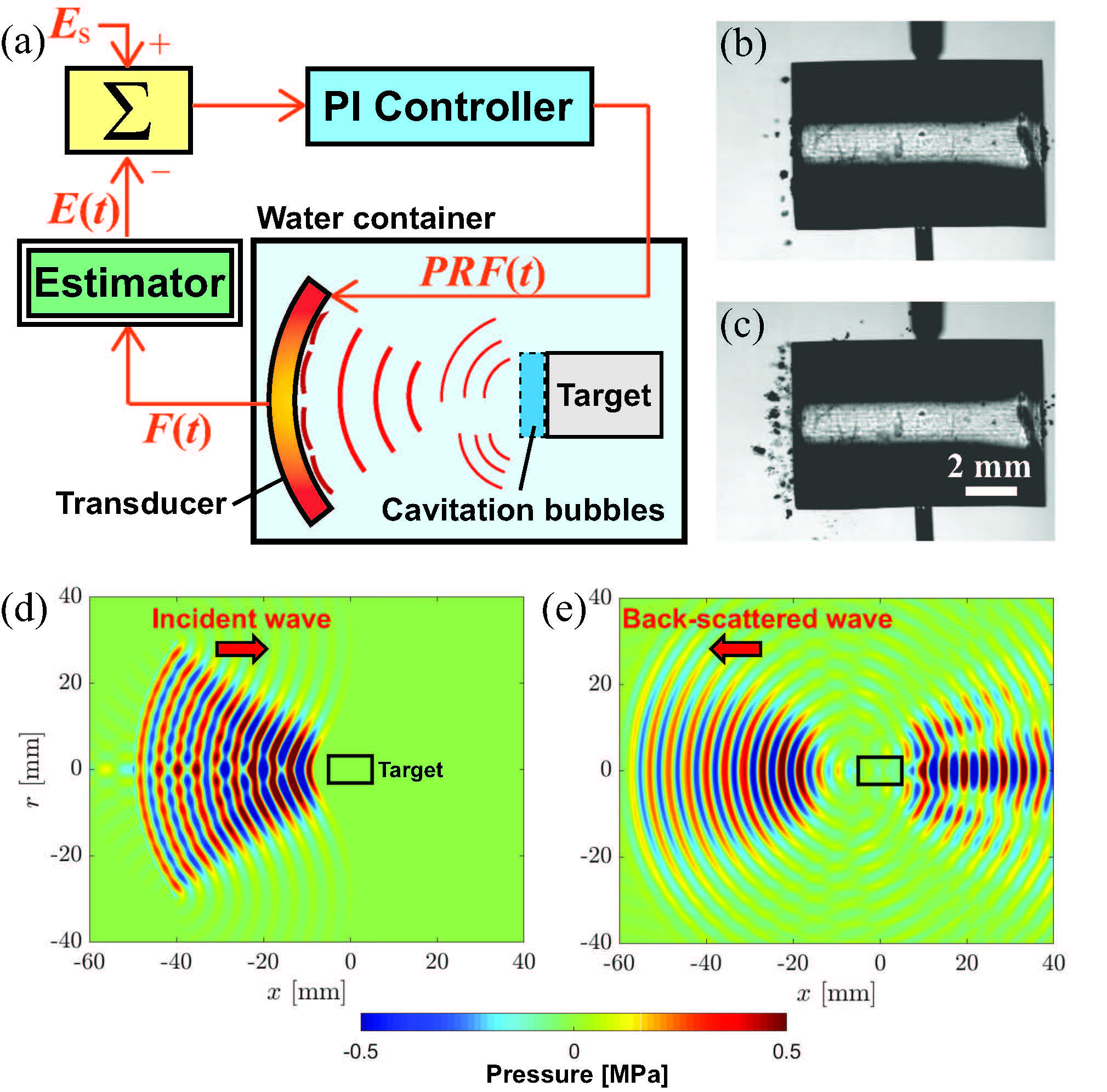}
\caption{(a) Schematic of the setup. A transducer generates US pulses focused on a cylindrical target, aligned on the acoustic axis, and measured the back-scattered acoustic waves, in a water container. A model-based feedback controller modulates PRF based on the estimation of the energy transmitted into the target, $E$. (b) Experimental high-speed images of the bubble clouds with a PRF of 10 Hz and (c) that with 100 Hz during a short-term exposure of US pulses. (d) Instantaneous pressure contour during the propagation of the incident US wave obtained from a representative numerical simulation. (e) That of the wave scattered by the target shielded by a bubbly layer.}
\label{f:f1} 
\end{figure}
Figure 1 shows the schematic of the control and the US setup.
In the setup, pulses of a US wave are generated by a transducer and focused on a cylindrical target made of epoxy resin, with the base diameter and the length of 6.25 and 10 mm. Each pulse contains 10 cycles of a sinusoidal wave packet with a frequency of 340 kHz and a peak focal amplitude of 7.0 MPa.
During the passage of the wave, a layer of cavitation bubbles is nucleated on the proximal surface of the target to cause energy shielding.
The acoustic waves scattered by the bubbles are measured using an array transducer.
Further technical details of this setup (without control) are described elsewhere \citep{Maeda18a}.
The controller varies the Pulse-Repetition-Frequency (PRF);
the intensities of both cavitation and the bubble-scattered acoustics are positively correlated with PRF (fig 1b,c); large PRF indicates short pulse interval. With a short interval, fewer bubbles tend to dissolve before the next pulse arrives, and cavitation is enhanced pulse-to-pulse, leading to the proliferation. For the dense cloud (fig 1c), the bubbles in the cloud most proximal to the transducer are excited more than the distal ones. A similar anisotropic structure was previously observed in isolated, spherical bubble clouds \citep{Maeda19}.
The measurement is used to estimate the energy that is transmitted across the bubble clouds into the target.
Based on the offset of the estimation from a set point, a proportional-integral (PI) controller varies PRF, but with a constant pressure amplitude, to modulate cavitation and to achieve the set point value.
Using the setup, we experimentally demonstrate real-time control of cavitation during $O$(100) s of pulsed US radiation and identify an empirical correlation between PRF and the bubble dynamics, and an optimal PRF to minimize the shielding by cavitation.

The acoustic signals measured at each array element are projected onto a correlation-based imaging functional $F$ \citep{Garnier09}, which is used as an estimator input.
$F$ is a normalized measure of the amplitude of the coherent acoustic scattering from the target region:
$F={I}/{I_0}$,
where
\begin{equation}
I=MAX_{\vector{z}}[\sum_{j,l}C_T(\tau(\vector{z},\vector{x}_j)+\tau(\vector{z},\vector{x}_l), \vector{x}_j, \vector{x}_l)],
\end{equation}
and $C_T$ is the cross-correlation of the signals:
\begin{equation}
C_T(\tau, \vector{x}_j, \vector{x}_l) = \frac{1}{T}\int^T_0u(t, \vector{x}_j)u(t+\tau, \vector{x}_l)dt,
\end{equation}
where $u(t,\vector{x}_j)$ is the signal at the $j$-th array at $\vector{x}_j$ at time $t$, $\vector{z}$ is the coordinate of the domain, and $T$ is the time horizon of the process. $\tau(\vector{z},\vector{x})$ is the acoustic travel time from $\vector{z}$ to $\vector{x}$.
$I_0$ is the reference value obtained with a case without bubbles.

The estimator outputs a scalar variable $E$, which is defined as the energy transmitted across bubbles into the target, normalized by the reference value obtained without bubbles.
This choice of the input and output parameters is based on a first-principle theory of the bubble cloud dynamics.
The kinetic energy of incompressible potential flow induced by oscillations of interacting spherical bubbles can be expressed as
\begin{equation}
K = 2\pi\rho_l\sum^N_{i=1} \left[R_i^3\dot{R}_i^2+\sum^N_{j\ne i}\frac{R_i^2R_j^2 \dot{R}_i \dot{R}_j}{r_{ij}}\right]+(H.O.T),
\end{equation}
where $R_i$, $\dot{R}_i$, and $r_{ij}$ are the radius and the radial velocity of bubble $i$, and distance between the centers of bubble $i$ and $j$, respectively.
For a cloud with $N\gg 1$, the kinetic energy can be approximated as
\begin{equation}
K \approx 2\pi\rho_lN<R^3\dot{R}^2>(1+B_d),
\end{equation}
where $B_d = N<R>/L$.
$L$ is the length-scale of the inter-bubble interaction.
$B_d$ is therefore a measure of the relative contribution of the coherence among bubbles.
Hydrodynamically, $B_d$ corresponds to the scale of the added-inertia of a bubble cloud.
For a spherical bubble cloud, $L\sim R_c$ and $B_d$ corresponds to the aforementioned dynamic interaction parameter.
For the present bubbly layer, $L\sim R_t$, where $R_t$ is the radius of the base of the target.
Since the bubble cloud is the only component that dynamically alters the energy state of the system, both $E$ and $F$ are expected to be scaled by $B_d:B_d = N<R>/R_t$, and thus these variables are correlated, regardless of $N$ and $<R>$.

Based on this physical insight, we conduct numerical experiments on a high-performance computer to obtain the quantitative correlation between $E$ and $F$.
A coupled Eulerian-Lagrangian method was employed for the simulation \citep{Maeda18b}. 
This numerical setup is designed to fully mimic the physical setup.
In the initial condition, Lagrangian bubble nuclei with a uniform radius of 10 $\mu$m are randomly distributed in the cylindrical region on the proximal base of the target.
Oscillations of the bubbles are tracked as solutions of the Keller-Miksis equation.
The US wave and bubble-scattered pressure waves are computed as solutions of the compressible Navier-Stokes equations on a structured Eulerian grid with a sufficiently high resolution.
Twenty cases were simulated during the passage and scattering of a single US pulse, with various values of the bubble's number density, $n$, and the thickness of the bubbly layer, $h$, within ranges of $0\le h\le1.0$ mm and $0\le n \le9.6$ mm$^{-3}$, respectively.

\begin{figure*}[t!]
\centering
     \includegraphics[width=115mm]{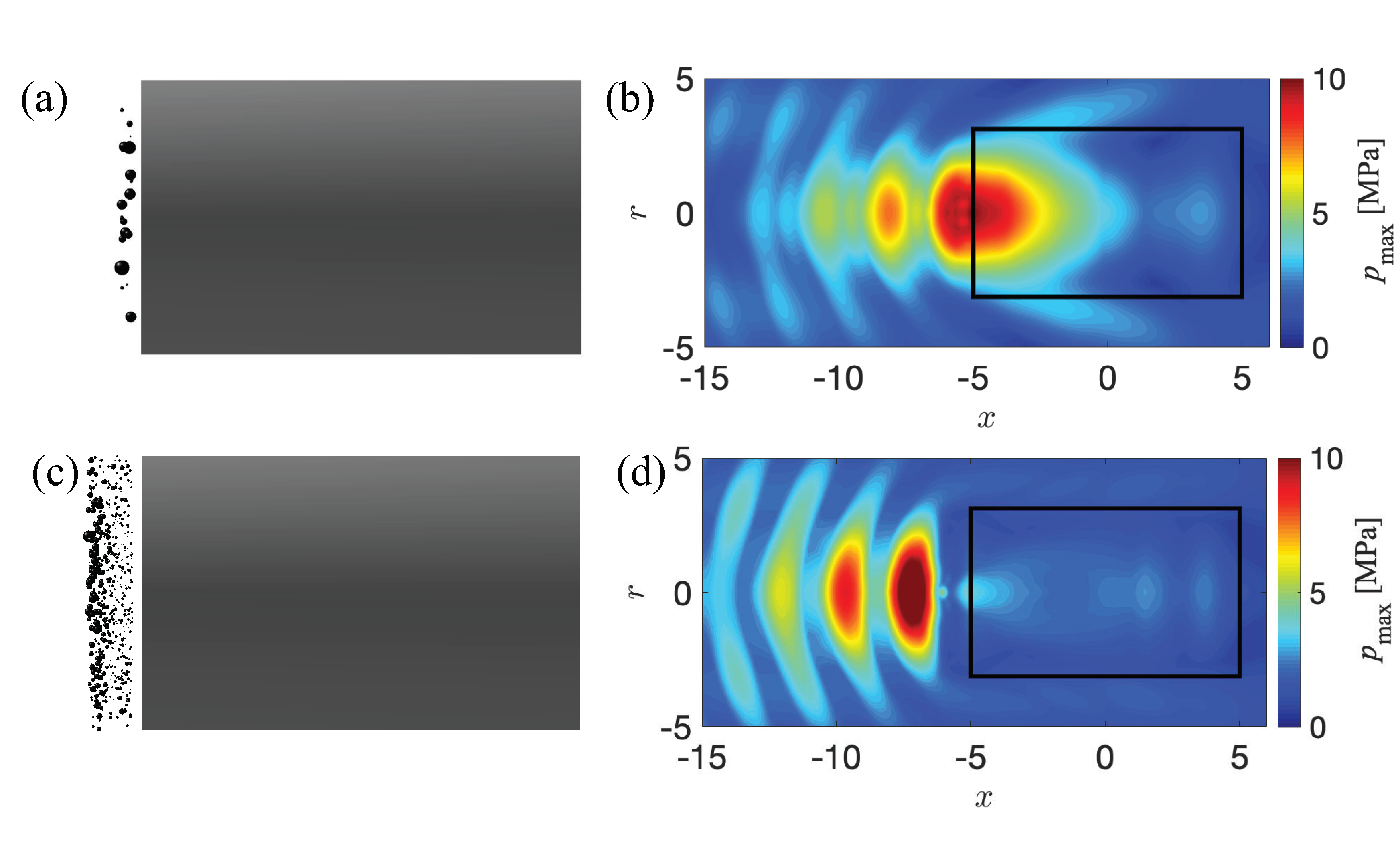}\label{<figure1>}
     \caption{(a,c) Side views of the bubbly layers and the target obtained from the simulations that model the low and high PRF cases during the passage of the US wave, with $(n,h)=(1.2,0.25)$ and $(9.6,1.0)$. An anisotropic structure, in that the proximal bubbles are more excited than the distal bubbles, is evident in the latter image. (b,d) Distributions of the maximum pressure during the course of the same simulations.}
     \label{f:snap}
\end{figure*}
\begin{figure}
\centering
     \subfloat[]{\includegraphics[width=55mm]{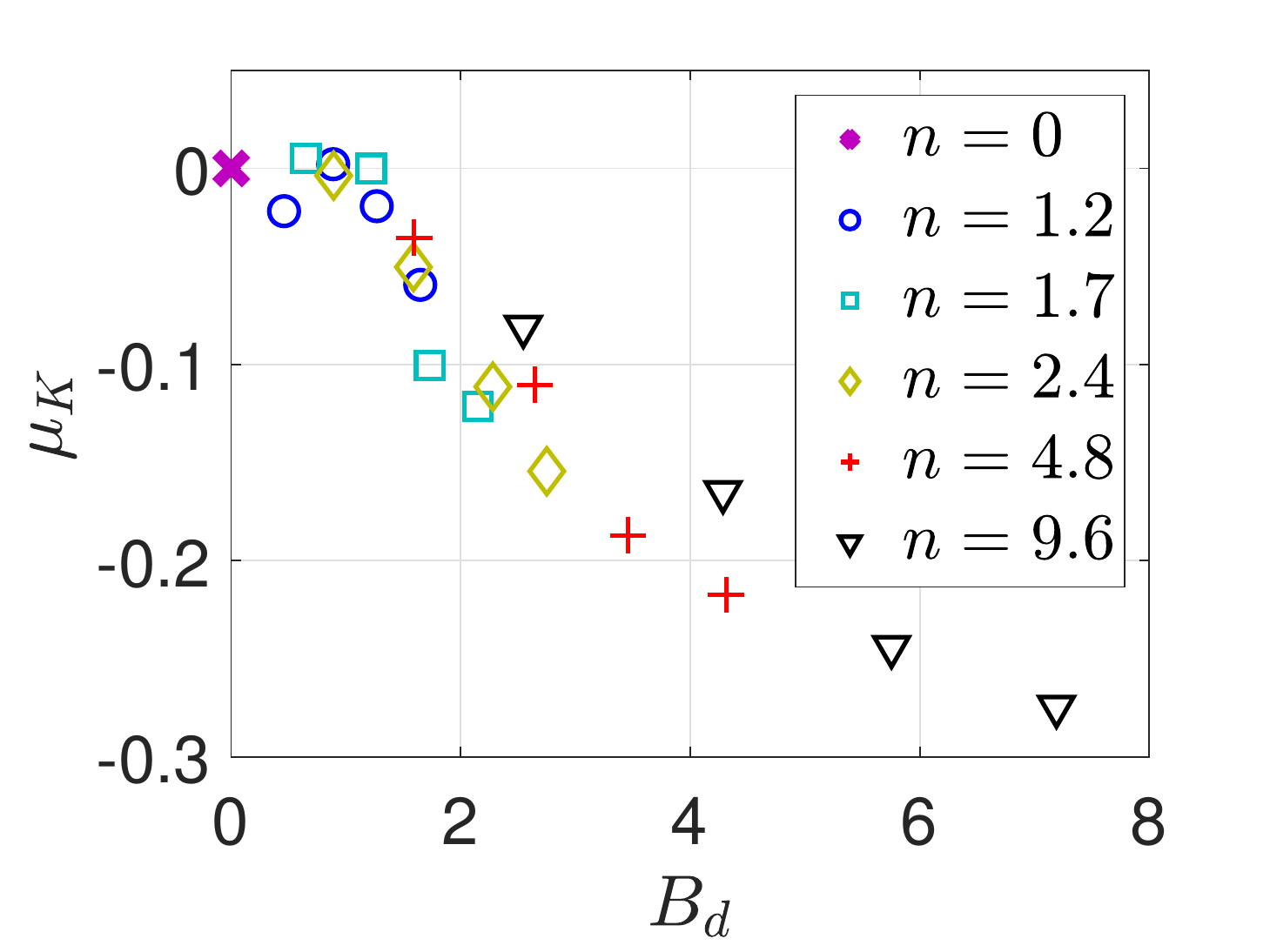}}
     \subfloat[]{\includegraphics[width=55mm]{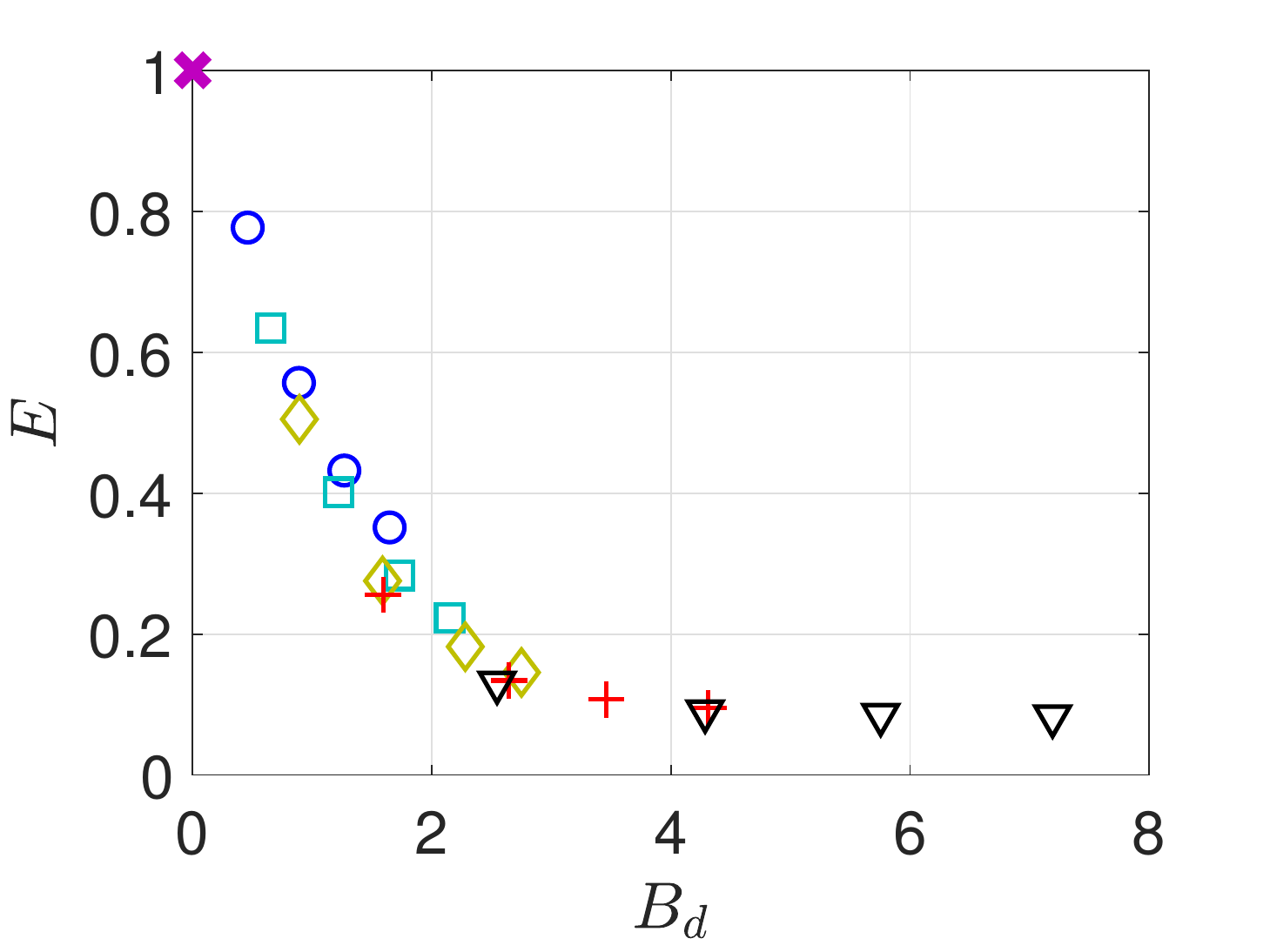}}\\
     \subfloat[]{\includegraphics[width=55mm]{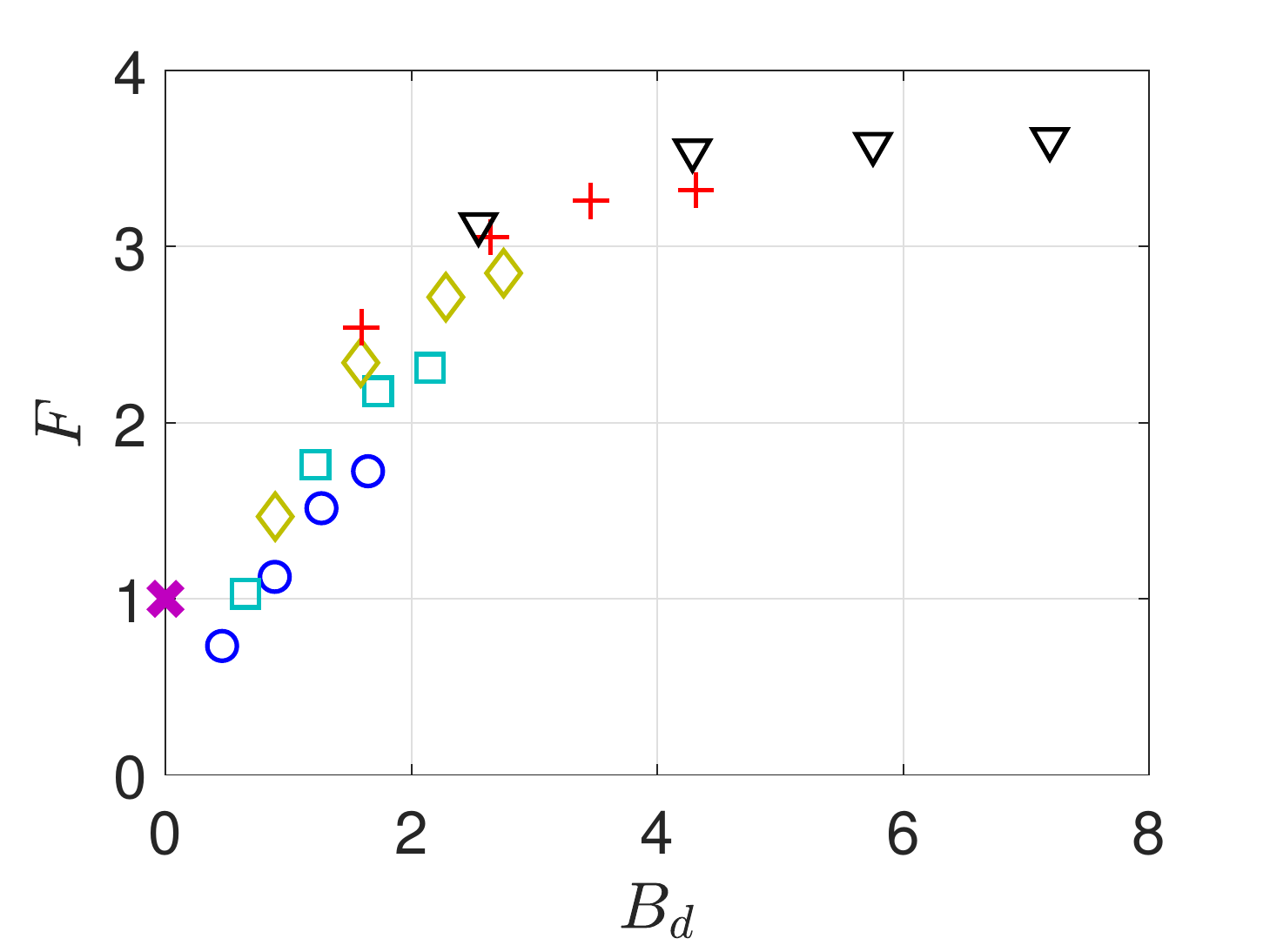}}
     \subfloat[]{\includegraphics[width=55mm]{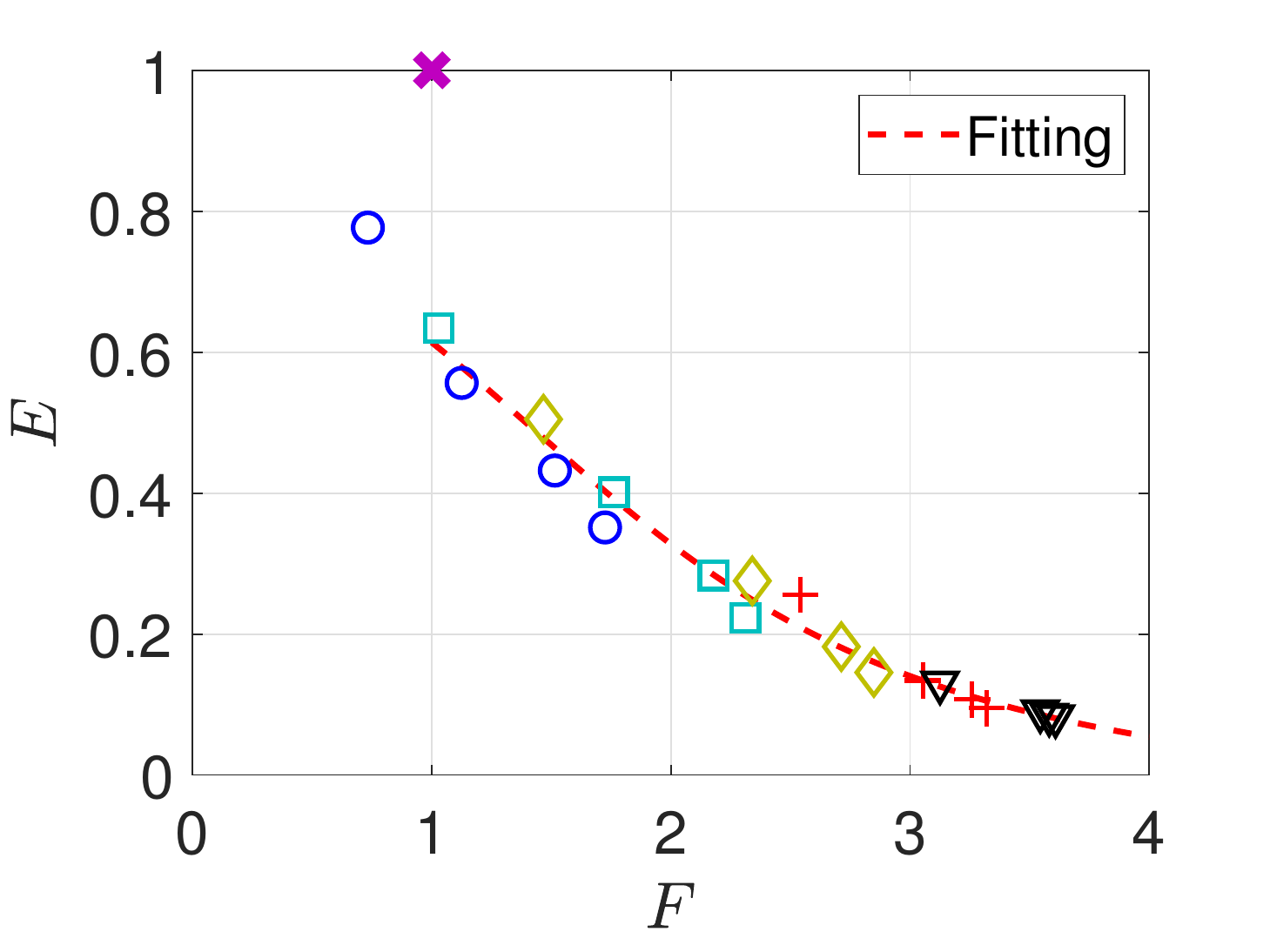}}
     \caption{(a-c) Correlations of $\mu_k$, $E$, and $F$ against $B_d$, respectively. The bubble's number density $n$ mm$^{-3}$ and the thickness of the bubble cloud $h:0<h<1.0$ mm are varied.
     (d) Correlation of $E$ and $F$. The fitting function serves as the state estimator shown in fig. 1a.}
     \label{f:sim}
\end{figure}

Figure \ref{f:snap}a and c show the snapshots of the bubbles during the passage of the US with $(n,h)=(1.2,0.25)$ and $(9.6,1.0)$, at the same instance as the experimental images (fig. \ref{f:f1}b,c). Figure \ref{f:sim}b and d show the contours of the maximum pressure on the cross-plane throughout the same simulations, $p_{max}$,  respectively.
The morphology of the numerical bubble clouds is similar to that in the experimental images. The anisotropic structure is clear in the dense numerical cloud. In the pressure contours, the region with a high maximum pressure ($p_{max}>6$ MPa) is widely distributed in the proximal interior of the target with the dilute bubbles, while the maximum pressure is nominally small ($p_{max}<4$ MPa) in the target with the dense cloud. These results indicate that the US wave penetrates into the target across the small bubble cloud, while a large portion of the wave energy is scattered by the large cloud. In the latter contour, clear vertical bands of high pressure are observed in the proximal liquid, which can be explained by the interference of the reflected and incoming parts of the wave.

To quantify the anisotropy, we use the normalized moment of kinetic energy, $\mu_K$, defined as
\begin{equation}
\mu_K=\frac{\sum^{N}_{i=1}K_i(x_i-x_c)}{\sum^{N}_{i=1}K_ih},
\end{equation}
where $K_i$ is the kinetic energy of incompressible liquid induced by the oscillations of $i$th bubble: $K_i=2\pi\rho R_i^3\dot{R}^2$, where $x_i$ is the coordinate of bubble $i$ along the acoustic axis, and $x_c$ is the center of the cloud: $x_c=\sum^N_{i=1}x_i/N$. Negative $\mu_k$ indicates the spatial bias of active bubbles in the proximal side of the cloud.
A similar moment was previously used to characterize the energy state of spherical bubble clouds \citep{Maeda19}.
Figure \ref{f:sim}a-c respectively show correlations of $\mu_K$, $E$, and $F$ against $B_d$, obtained from the simulation.
Data points are collapsed well in all plots. $\mu_K$ and $E$ decrease, while $F$ increases with increasing $B_d$.
$E$ and $F$ become invariant for $B_d>4$, indicating that the energy shielding is saturated in this regime.
Figure \ref{f:sim}d shows correlations between $E$ and $F$. The data points are well collapsed on a single curve. Note that the $E-F$ correlation is not guaranteed to be linear due to the non-linear dynamics of bubbles. The numerical simulation is therefore critical to obtain this correlation. Notice also that the correlation is non-monotonic for $F<1$; the reference state without bubbles ($n=0$) is placed at $(F,E)=(1,1)$. This anomaly can be explained by the breakdown of the scaling with $B_d$, for small $N$.

These analyses indicate that the bubble dynamics are dictated by the interaction parameter, and the acoustic wave and the energy transmission are monotonically correlated, as predicted by the theory.
With the increase in this parameter, the anisotropy and the scattering are enhanced, while the energy transmission decreases.
From a macro-scopic point of view, the greatest portion of the acoustic energy is scattered by only the surface bubbles when clouds are thick and/or dense, while otherwise a large portion of the acoustic energy is transmitted and all bubbles oscillate in a similar manner regardless of their locations.
We apply nonlinear regression to data in fig. 3d to obtain a fitting function. This function serves as the estimator that uniquely outputs $E$ against a real-time input of $F$, for $F>1$.

\begin{figure}[h!]
     \centering
     \subfloat[]{\includegraphics[width=100mm]{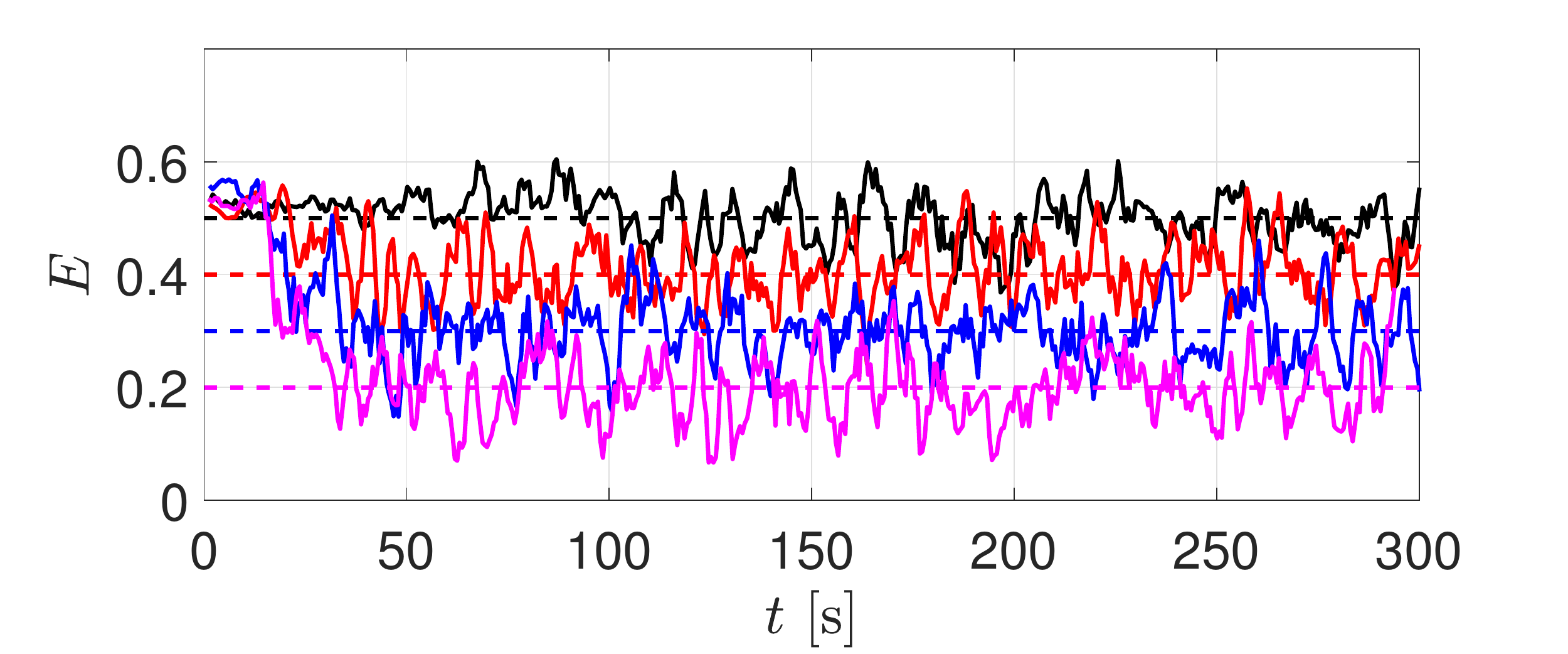}}\\
     \subfloat[]{\includegraphics[width=100mm]{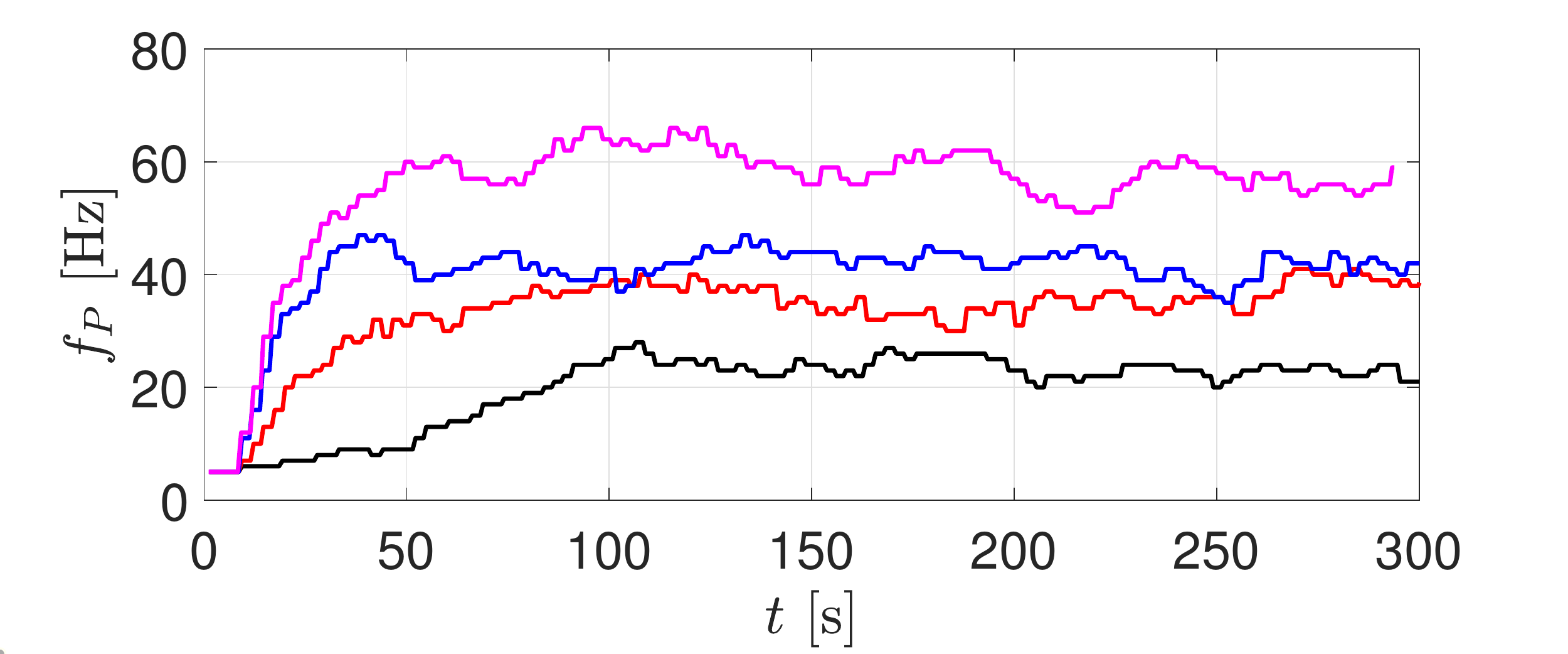}}
     \caption{Evolution of (a) $E$ and (b) PRF during feedback control with set point values of $E_s=[0.2,0.3,0.4,0.5]$. Dotted lines in (a) denote set point values.}
     \label{f:evol}
\end{figure}
Figure \ref{f:evol}a and b respectively show evolution of the energy transmission outputted by the estimator and that of PRF outputted by the controller during US radiation, with four distinct values of the set point: $E_s=[0.2,0.3,0.4,0.5]$.
The feedback rate was $f_P/5$ Hz, where $f_P$ is PRF, and $K_P = 20$ and $K_I=20$ were used in the controller.
At around $t = 100$ s, $E$ reaches its steady state and then oscillates around a corresponding set point value, for all cases.
Similarly, PRF evolves with constant oscillations around its stationary values for $t>100$ s, in the range of $20<f_P<70$ Hz.
Note that without control, cavitation bubbles intermittently proliferate over the timescale considered in the present study, and this phenomena is not even reproducible trial by trial.
Although it is known that the increase in PRF can enhance the intensity of acoustic cavitation \citep{Pishchalnikov11,Maxwell11,Frank15}, to our knowledge there exists no quantitative measure to uniquely specify the bubble dynamics given PRF, due to this extreme nonstationarity.
With the real-time control, cavitation can, for the first time, achieve quasi-equilibrium over pulses.
We are thus motivated to correlate PRF (controller output) and the transmitted energy (estimator output) in these states.

\begin{figure}
     \centering
     \subfloat[]{\includegraphics[width=55mm]{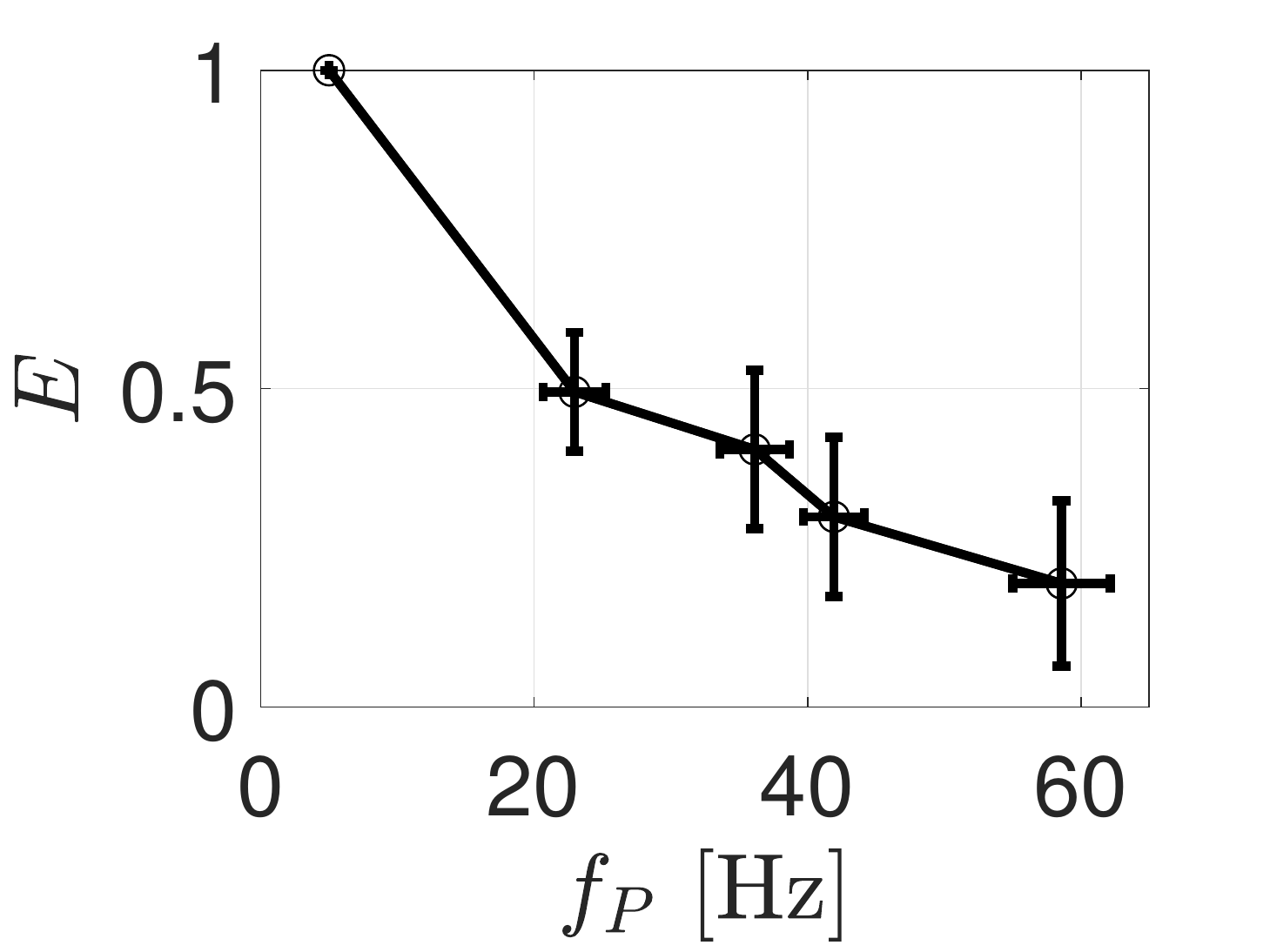}}
     \subfloat[]{\includegraphics[width=55mm]{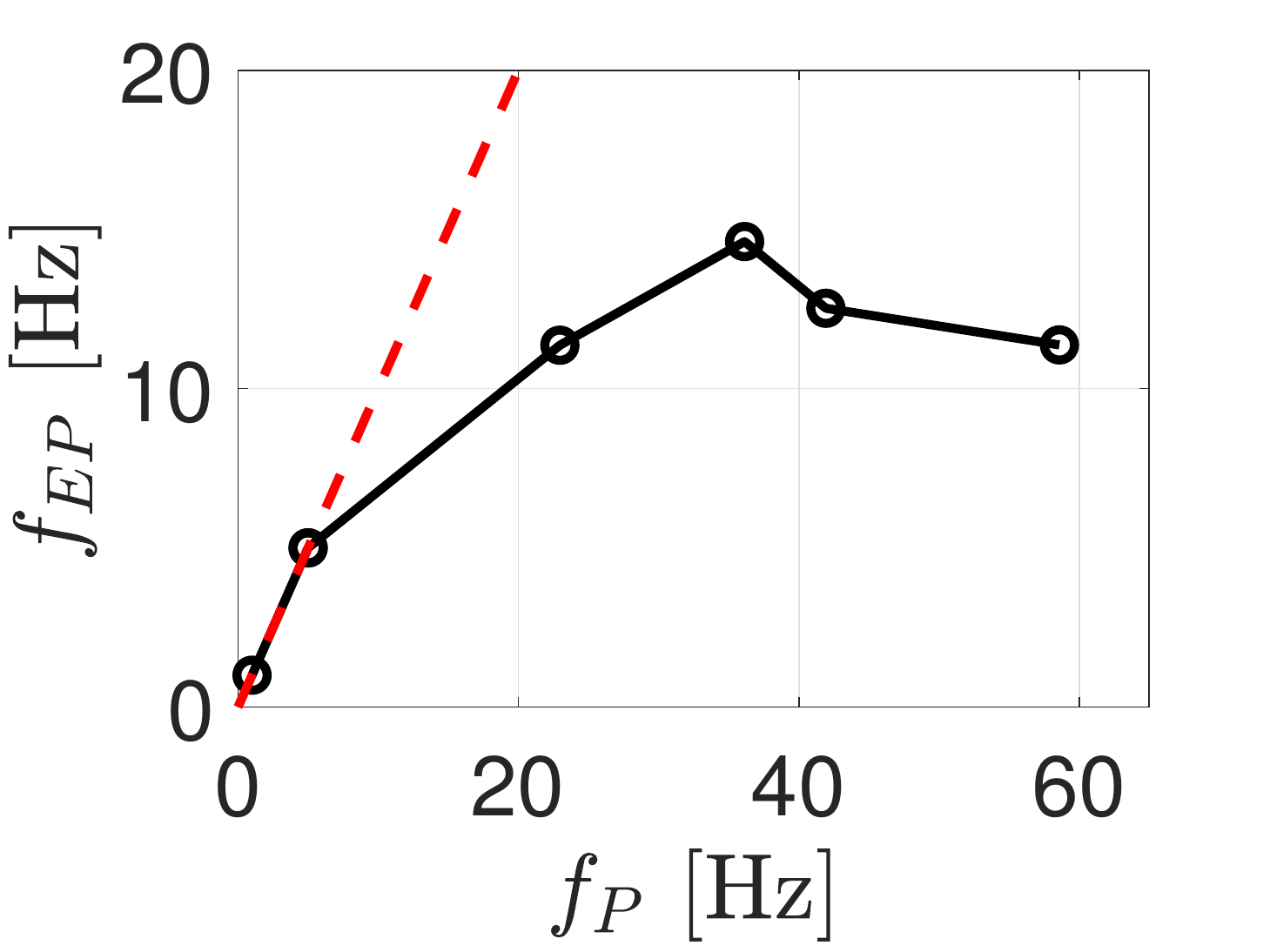}}
     \caption{(a)The energy transmitted into the target, $E$, as a function of actual PRF. Both variables are averaged over the period of the feedback control during $100<t<300$ s. (b) The effective PRF as a function of actual PRF, in terms of the mean values during feedback control. The dotted line denotes linear reference with $E=1$.}
     \label{f:prf}
\end{figure}
Figure \ref{f:prf}a shows the correlation. For $f_P<5$ Hz, we did not observe bubbles and $F<1$. In this range, we set $E$ to unity. For $f_P>5$ Hz, the energy monotonically decreases, indicating the enhancement of cavitation with increasing $f_P$. In order to further characterize the correlation between PRF and the energy transmission, we define the effective rate of energy delivery to the target excluding the portion scattered by the bubbly layer, {\it Effective-PRF}: $f_{EP}=E f_P$.
Interestingly, $f_{EP}$ has a convex profile and takes a peak value of 14.6 Hz at $f_{PC}=35$ Hz (Fig. 5b), which can be interpreted as the optimal PRF for energy transmission.
For $f_P<5$ Hz, $E=1$ and $f_P=f_{EP}$ there is no effect of cavitation.
For $f_P<5$ Hz,
\begin{equation}
\frac{\partial f_{EP}}{\partial f_P} =E+f_P\frac{\partial E}{\partial f_P}
\approx E+\gamma f_P,
\label{eqn:g}
\end{equation}
\iffalse
For $f_P<5$ Hz, $E=1$ and $f_P=f_{EP}$ ($\partial f_{EP}/\partial f_P=0$); there is no effect of cavitation.
For $f_P>5$ Hz,
\begin{equation}
\frac{\partial f_{EP}}{\partial f_P}\approx E+\gamma f_P,
\label{eqn:g}
\end{equation}
\fi
where $\gamma$ is a constant that approximates the slope of $E$:
$\partial E/\partial f_P\approx \gamma<0$.
$\partial^2 f_{EP}/\partial f_P^2\approx 2\gamma<0$, indicating the convexity of $f_{EP}$.
For $5<f_P<f_{PC}$, both $E$ and $\gamma f_P$ decrease with increasing $f_P$, while it remains that $\partial f_{EP}/\partial f_P>0$; the rate of increase in $f_P$ is more dominant than that of the energy loss due to cavitation, respectively represented by $E$ and $\gamma f_P$ on the r.h.s. of (\ref{eqn:g}). Then, at $f_P\approx f_{PC}$, $\partial f_{EP} /\partial f_P=0$ and $f_{P}$ reaches the maximum value.
For $f_P>f_{PC}$, $\partial f_{EP} /\partial f_P<0$; $f_{EP}$ decreases with $f_{P}$ since $\gamma f_P$ is dominant.
At even higher $f_{P}$, $E<0.2$; cavitation and the energy shielding become further enhanced. In this regime, the present estimator tends to become inaccurate; the state of the bubble cannot be uniquely identified by $E$, as seen in the saturation of $E$ against $B_d$.

In conclusion, we have designed a framework for model-based closed-loop control of cavitation through acoustic feedback by using a data-driven state estimator.
In our demonstration using a US system, set-point control of PRF modulated cloud cavitation near a solid target as designated, and identified an optimal PRF that can minimize the cavitation-induced energy shielding.
Although the nucleation and dissolution of bubbles are not explicitly modeled, the control system can implicitly maintain and quantify the balance between those phenomena in dynamic equilibrium, through correlating estimation and control.
The framework can admit an arbitrary choice of control parameter for systems that rupture liquids using different energy source other than US.
The quality of state estimation can be improved by enhancing the precision of numerical data and acoustic measurements; the framework can be scaled up with computational power and acoustic sensing.
These flexibility and scalability may be of critical use for characterizing and optimizing cavitation in various regimes and in applications.
\newline
\newline
K.M. acknowledges NIH under grants P01-DK043881 and the Extreme Science and Engineering Discovery Environment (XSEDE), which is supported by NSF under grant TG-CTS190009.
A.D.M acknowledges NIH under grants P01-DK043881 and K01-DK104854.

%\section*{Citations}
\bibliographystyle{revtex}
%\bibliography{sample}

%\bibliographystyle{model1-num-names}
%% Authors are advised to submit their bibtex database files. They are
%% requested to list a bibtex style file in the manuscript if they do
%% not want to use model1-num-names.bst.

%% References without bibTeX database:

% \begin{thebibliography}{00}

%% \bibitem must have the following form:
%%   \bibitem{key}...
%%

% \bibitem{}

% \end{thebibliography}
%\providecommand{\noopsort}[1]{}\providecommand{\singleletter}[1]{#1}%

\end{document}